\documentclass[12pt,preprint]{aastex}

\def\wig#1{\mathrel{\hbox{\hbox to 0pt{%
          \lower.5ex\hbox{$\sim$}\hss}\raise.4ex\hbox{$#1$}}}}

\shorttitle{Structure of Giant Planets}
\shortauthors{Fortney}

\newcommand{\mj}{$M_{\mathrm{J}}$}
\newcommand{\rj}{$R_{\mathrm{J}}$}
\newcommand{\me}{$M_{\oplus}$}

\newcommand{\hd}{HD 209458b} 
\newcommand{\hh}{HD 149026b}

\slugcomment{\bf}
\slugcomment{Proceedings of HEDLA 6\\Accepted to Astrophysics \& Space Science, August 1, 2006}

\begin{document}

\title{The Structure of Jupiter, Saturn, and Exoplanets:\\Key Questions for High-Pressure Experiments}

\author{Jonathan J. Fortney\\ \emph{Space Science and Astrobiology Division, NASA Ames Research Center, MS 245-3,\\Moffett Field, CA 94035;} jfortney@arc.nasa.gov}

\begin{abstract}

We give an overview of our current understanding of the structure of gas giant planets, from Jupiter and Saturn to extrasolar giant planets.  We focus on addressing what high-pressure laboratory experiments on hydrogen and helium can help to elucidate about the structure of these planets.

\end{abstract}



\section{Introduction}
In order to understand the formation of giant planets, and hence, the formation of planetary systems, we must be able to determine the interior structure and composition of giant planets.  Jupiter and Saturn, our solar system's gas giants, combine to make up 92\% of the planetary mass of our solar system.  Interestingly, knowledge of only a few key quantities allows us to gain important insight into their interior structure.  The equation of state of hydrogen, together with measurements of the mass, radius, and oblateness of Jupiter and Saturn is sufficient to show that these planets are hydrogen-helium rich objects with a composition similar to that of the Sun \citep{Demarcus58}.  Furthermore, estimates of the transport coefficients of dense metallic hydrogen and the observation that Jupiter emits more infrared radiation than it absorbs from the Sun \citep{Low66}, is sufficient to show that gas giant planet interiors are warm, fluid, and convective, not cold and solid \citep{Hubbard68}.  It has also been clear for some time that the composition of Jupiter and Saturn is not exactly like that of the Sun---these planets are enhanced in ``heavy elements'' (atoms more massive than helium), compared to the Sun \citep{Podolak74}.  An understanding of how these planets attained these heavy elements, and their relative ratios, can give us a wealth of information on planetary formation and the state of the solar nebula.

Looking beyond Jupiter and Saturn, we now have 200 extrasolar giant planets (EGPs) that have been found to orbit other stars.  A subclass of these planets are the ``hot Jupiters'' that orbit their parent stars at around 0.05 AU.  To date, ten planets (with masses from 0.36 to 1.5 \mj) have been seen to transit their parents stars.  All of these objects are hot Jupiters, with orbital periods of only a few days \citep[see][]{Charb06}.  These transiting planets are important because we can measure their masses \emph{and} radii, thereby allowing us access to information on their interior structure \citep{Guillot05}.  While our understanding of the interiors of these planets will never be as detailed as that for Jupiter and Saturn, we will eventually have a very large sample of these transiting objects at various masses, compositions, and orbital distances, which will allow for an understanding of the mass--radius relation for giant planets under a variety of conditions.

By far the most important physical input into giant planet structural models is the equation of state (EOS) of hydrogen.  The decade of pressure that is most important for understanding the interiors of giant planets is 1-10 Mbar (100-1000 Gpa) \citep{Saumon04}.  In the past decade experiments have been able to probe into the lower end of this pressure range \citep{Weir96, Collins98, Knudson01, Boriskov05}.  In this paper, instead of focusing on equation of state physics we will focus on key questions for understanding the structure and composition of giant planets.  As we discuss giant planet interiors we will investigate how high pressure laboratory experiments have and will continue to allow us to better answer these questions.

\section{Key Questions}
\subsection{Are planetary atmospheric abundances representative\\of the entire H/He envelope?}
This question is directly related to whether hydrogen's molecular-to-metallic transition is continuous or first-order.  Whether or not hydrogen's transition to a metal in the fluid state is first order has always been an open issue.  The importance of this question to giant planets cannot be overstated.  If the transition is first order (a so-called ``plasma phase transition,'' or PPT) then there will be an impenetrable barrier to convection within the planet and there must also be several discontinuities at this transition.  One is a discontinuity in entropy \citep{SS77b, SC92}.  In the 1970s, W.~B.~Hubbard discussed that, for a fully convective and adiabatic giant planet, a measurement of of the specific entropy in the convective atmosphere would essentially allow us to understand the run of temperature vs.~pressure for the entire planet, as all regions would share this specific entropy \citep[see][]{Hubbard73}.  However, if a PPT exists, this will not be true \citep{Chabrier92}.

Another discontinuity at the PPT would be in chemical composition, due to the Gibbs phase rule.  Modern structural models of Jupiter and Saturn aim to constrain the bulk abundance and distribution of heavy elements in the interiors of these planets.  We would like to understand what fraction of the heavy elements are distributed throughout the H/He envelope, and what fraction are in a central core.  See \citet{Guillot99} and \citet{Saumon04,Saumon05} for recent computations of the interior structure of Jupiter and Saturn.  The main constraints on these models are planetary mass, radius, rotation period, and gravity field.  Additional constraints would be most welcome.  One potentially important constraint would be atmospheric abundances derived from entry probes or spectra.  If it could be clearly shown that the molecular-to-metallic transition is indeed continuous, then mixing ratios of chemical species in the atmosphere should be representative of the entire H/He envelope, as the entire envelope should be well-mixed due to efficient convection.  This could constrain the amount of heavy elements in the H/He envelope and allow for a much more precise determination of the core mass and bulk heavy element abundance.  For Jupiter, the \emph{Galileo Entry Probe} has measured the abundances of the important species methane and ammonia \citep{Atreya03}.  However, the abundance of water, presumably the most abundant species after helium, is still highly uncertain.

Perhaps the clearest indication of the physical state of hydrogen in the molecular-to-metallic transition region ($\sim$1-5 Mbar) would be a measurement of the hydrogen's conductivity.  To date, \citet{Weir96} and \citet{Nellis99} have measured the conductivity of hydrogen using a reverberation shock technique up to 1.8 Mbar (180 Gpa).  They found a four order of magnitude increase in conductivity from 0.93 to 1.4 Mbar that plateaued between 1.4 and 1.8 Mbar at a conductivity consistent with that of the minimum conductivity of a metal.  These measurements appear to indicate that hydrogen's transition to a metallic state is indeed continuous (at least at their measured temperature of 2600 K).  However, the measured conductivity is still over an order of magnitude less than that expected for a fully ionized hydrogen plasma \citep{Hubbard02}, so these measurements cannot be considered a definitive refutation of a PPT.  Another open question is how the presence of neutral atomic helium (10\% by number in a solar composition mixture) may affect this transition.

\subsection{Heavy Elements:  How much and where are they?}
The pressure-density relation of hydrogen is the single most important input in giant planet structural models.  All things being equal, the more compressible hydrogen is, the smaller a planet will be at a given mass and composition.  This has a direct bearing on model-derived constraints on the amount of heavy elements within a planet's interior.  \citet{Saumon04} computed detailed interior models for Jupiter and Saturn that were consistent with all available observational constraints.  They found that Jupiter models that used EOSs consistent with the 6-fold limiting deuterium compression data of \citet{Collins98} lead to core sizes of 0-10 \me,~with total heavy element abundances (envelope plus core) of 10-25 \me.  Models computed using EOSs consistent with the harder 4.3-fold limiting compression of \citet{Knudson01,Knudson04} and \citet{Boriskov05} led to smaller cores sizes (0-3 \me) but larger heavy elements abundances (25-35 \me).  Since other experiments have not been able to replicate the soft \citet{Collins98} data, and the data of Knudson et al.~and Boriskov et al.~agree quite well while using different experimental setups, these harder EOS data sets are currently viewed by many as the most reliable.  \citep[For recent reviews, see][]{Nellis05,Nellis06}.  Tests of the hydrogen or deuterium EOS off of the single-shock Hugoniot, perhaps at pressures of up to a few Mbar, but temperatures below 10$^4$ K, would be most valuable.  For helium, our second most important constituent,  new EOS data are sorely needed.  No helium EOS data have been published since \citet{Nellis84}, and this data set only reached a maximum pressure of 560 kbar (56 Gpa).

\begin{figure}
\plotone{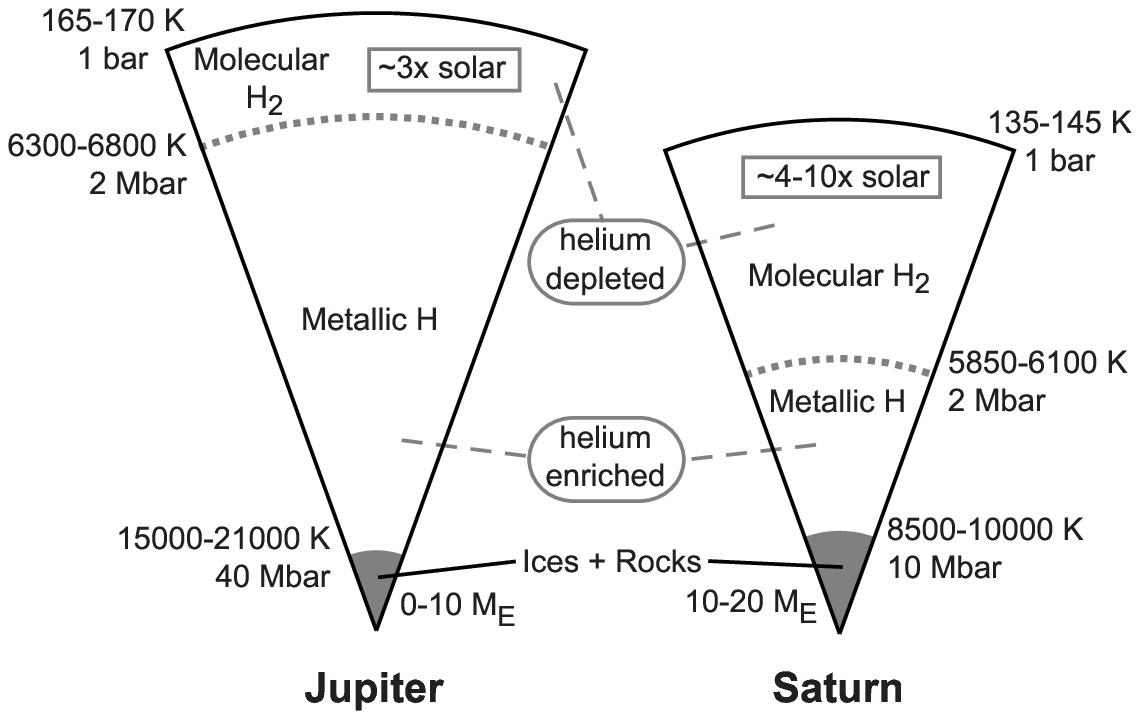}
\caption{Schematic interior structure of Jupiter and Saturn.  Pressures and temperature are marked at 1 bar (100 kpa, visible atmosphere), 2 Mbar (200 Gpa, near the molecular-to-metallic transition of hydrogen), and at the top of the heavy element core.  Temperatures are especially uncertain, and are taken from \citet{Guillot05}.  Approximate atmospheric abundances for ``metals'' (relative to solar) are shown within the grey box, in the molecular H$_2$ region.  Possible core masses, in \me~(labeled as ``M$_{\rm E}$'') are shown as well \citep{Saumon04}.
\label{figure:JS}}
\end{figure}

In \mbox{Figure~\ref{figure:JS}} we show schematic interior structures of Jupiter and Saturn.  We show pressures and temperatures at three locations: the visible atmosphere (1 bar), near the molecular-to-metallic transition of hydrogen (2 Mbar), and at the top of the heavy element core of each planet.  Atmospheric elemental abundances, as determined by the \emph{Galileo Entry Probe} for Jupiter and by spectroscopy for Saturn, are shown within a grey box \citep{Atreya03}.  These abundances should at least be representative of the entire molecular H$_2$ region.  If a PPT does not exist, these abundances should be representative of the entire H/He envelope.  In both planets, the molecular H$_2$ region is depleted in helium relative to protosolar abundances \citep{vonzahn98,CG00} indicating sedimentation of helium into metallic H layers.  Recent evolutionary models for Saturn indicate this helium may rain through the metallic H region and form a layer on top of the core \citep{FH03}.

\subsection{What are the temperatures in the deep interiors of Jupiter and Saturn?}
While the interior pressure-density relation sets the structure of the planet, it is the pressure-temperature relation that determines the thermal evolution.  The temperature of the deep interior sets the heat content of the planet.  The higher the temperatures in the planet's interior, the longer it will take to cool to a given luminosity.  This has been investigated recently by \citet{Saumon04} for Jupiter.  They computed evolution models of Jupiter using several different hydrogen EOSs that span the range of data obtained from LLNL laser \citep{Collins98} and Sandia Z \citep{Knudson04} data.  These different EOSs predict temperatures than can differ by as much as 30\% at 1 Mbar.  They find that Jupiter models cool to the planet's known luminosity in $\sim$3 to 5.5 Gyr using these various EOSs.  This 2.5 Gyr uncertainty is rather significant.

The atmospheres of Jupiter and Saturn are both depleted in helium relative to protosolar composition \citep{Atreya03}.  This observation, together with theoretical work indicating that helium has a limited solubility in metallic hydrogen at planetary interior temperatures of $\sim$10$^4$ K \citep{Stevenson75,HDW,Pfaff}, indicates helium is phase separating from hydrogen and being lost to deeper layers in each planet.  The evolution of Saturn, and perhaps Jupiter, must be able to accommodate the substantial additional energy source due to differentiation within the planet.   This ``helium rain,'' if present, has been shown to be the dominant energy source for several-Gyr-old giant planets \citep{SS77b, FH03, FH04}.  In order to understand to what degree helium phase separation has progressed in Jupiter and Saturn, and how far down into the planet the helium has rained to, we must understand the deep interior temperature of these planets.

To date, temperature measurements have been published by \citet{Holmes95} and \citet{Collins01}.  These experiments were performed using gas gun and laser apparatuses, respectively.  Both found temperatures generally lower than most calculated hydrogen EOSs, which if indeed correct, would lead to shorter cooling timescales for giant planets.  This faster cooling would more easily accommodate the additional energy source due to helium rain.  Additional data, especially at the high pressures and ``cool'' temperatures of planetary interest (off of the single-shock Hugoniot) would be of great interest.

\subsection{Do all giant planets possess heavy element enrichments?}
\begin{figure}
\plotone{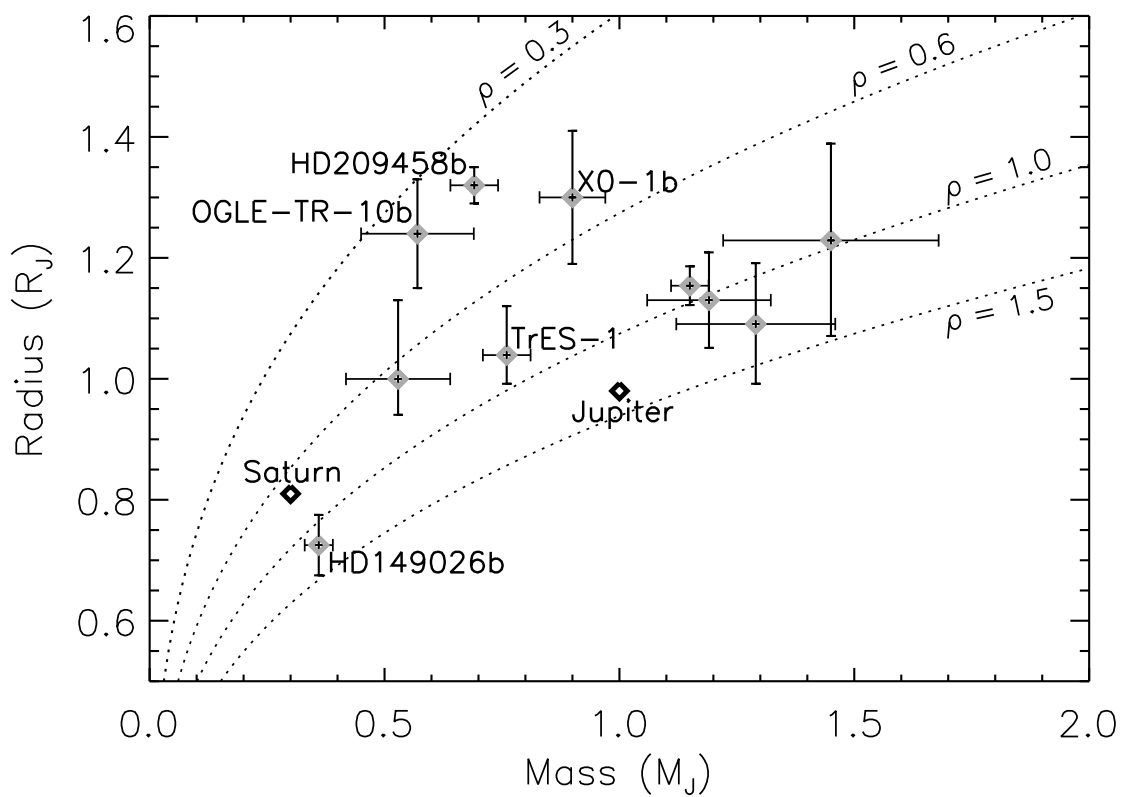}
\caption{Radius and mass of Jupiter, Saturn, and the 10 known transiting hot Jupiters, as of April 2006.  See \citet{Charb06} and references therein.  One \rj~is 71492 km, Jupiter's equatorial radius at $P$=1 bar.  Curves of constant density (in g cm$^{-3}$) are overplotted with a dotted line.  Data are taken from \citet{Charb06} and \citet{McCullough06}.
\label{figure:RvsM}}
\end{figure}

If we are to understand giant planets as a class of astronomical objects, we must understand how similar other giant planets are to Jupiter and Saturn.  The mass-radius relation of exoplanets allows us, in principle, to understand if these planets have heavy element enrichments that are similar to Jupiter and Saturn. \mbox{Figure~\ref{figure:RvsM}} shows the mass and radius of Jupiter, Saturn, and the 10 known transiting hot Jupiters.  It is interesting to note while Jupiter and Saturn differ in mass by a factor of 3.3, their radii only differ by 18\%.  However, while the hot Jupiters differ in mass by a similar factor (of 4) they differ in radius by a factor of 2.  This large spread is presumably due to large difference in the interior heavy element abundances of these planets \citep{Fortney06, Guillot05,Guillot06}.  Giant planets under intense stellar irradiation cool and contract more slowly that those far from their parent stars, so radii larger than 1 \rj~are expected \citep{Guillot96}.

Planet \hh, with a radius of only 0.73 \rj, must be on the order of 2/3 heavy elements by mass to explain its small radius \citep{Sato05, Fortney06}.  Its parent star has a metallicity 2.3$\times$ that of the Sun, so this may point to a connection between stellar and planetary abundances.  However, the determination of planetary core sizes appears to be complicated by the need for an additional interior energy source (yet to be definitely identified) for planet \hd, and perhaps also OGLE-Tr-10b \citep{Bodenheimer01, Guillot02, Winn05}.  These planets have radii that are too large to be explained by conventional cooling/contraction models \citep{Chabrier04,Laughlin05}.  Therefore, the spread in \mbox{Figure~\ref{figure:RvsM}} is likely due to a combination of differing magnitudes of this interior energy source and heavy element abundances, which adds significant complications to this picture.  \citet{Guillot06} have recently proposed a correlation between the heavy element abundances in transiting planets and the metallicity of the planets' host stars, assuming an additional energy source that scales linearly with the incident stellar flux absorbed by the planets.

\begin{figure}
\plotone{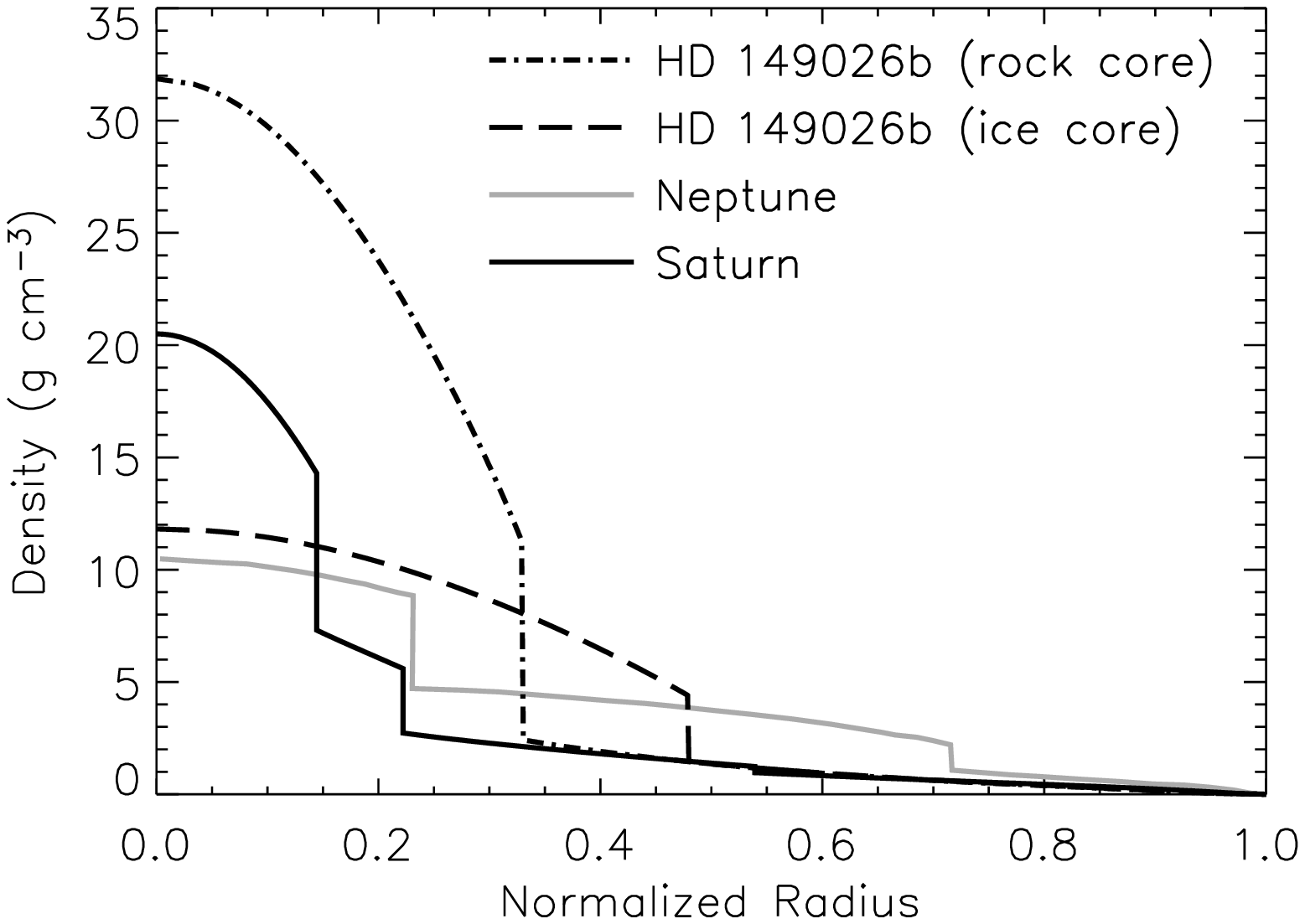}
\caption{Interior density as a function of normalized radius for two possible models for \hh~compared with Neptune and Saturn.  All planet models have been normalized to the radius at which $P$=1 bar.  The Neptune profile is from \citet{Podolak95} and the Saturn profile is from \citet{Guillot99}.  The Saturn and Neptune models have a two-layer core of ice overlying rock.  The two profiles of \hh~assume a metallicity of 3 times solar in the H/He envelope and a core made entirely of either ice or rock.
\label{figure:int}}
\end{figure}

In \mbox{Figure~\ref{figure:int}} we show a first look at comparative interior structure of the core-dominated planet \hh, Saturn, and Neptune.  The figure shows the current interior density distribution as a function of normalized radius for two \hh~models from \citet{Fortney06} compared to interior models of Saturn \citep{Guillot99} and Neptune \citep{Podolak95}.  The Saturn and Neptune models both have two-layer cores of rock overlain by ice.  The ratio of ice to rock in these cores is based more on cosmogonical arguments than on physical evidence.  The interior structure of \hh~may be a hybrid of the ice giants and gas giants.  Uranus and Neptune are $\sim$90\% heavy elements, while Saturn is $\sim$25\% and Jupiter $\wig<$10\% \citep{Saumon04}.  Although \hh~is more massive than Saturn, it has a bulk mass fraction of heavy elements (50-80\%) more similar to that of the solar system's ice giants.  Clearly, the field of exoplanets is allowing us to study and understand planets unlike any we have in our solar system.

\section{The Future}

The path towards a better understanding of the structure of giant planets seems clear.  Along with additional laboratory work at high irradiance laser, Z-pinch, and other facilities, space missions will also allow us better insight into giant planets.  For Saturn, NASA's \emph{Cassini} spacecraft will allow us to place better constraints on Saturn's gravity field.  For Jupiter, NASA's \emph{Juno} mission, still scheduled to launch in 2010, will map the planet's gravity field at high precision and to high order, and will derive the abundances of water vapor and ammonia in the planet's atmosphere below their respective cloud layers.  For extrasolar planets, the European \emph{COROT} and NASA \emph{Kepler} missions will allow us to detect potentially hundreds of additional transiting planets.  The scientific gain from all of these missions is directly dependent on our understanding of hydrogen and helium at high pressure.  Experiments in the future should focus on the following issues:
\begin{itemize}
\item{Is the fluid molecular-to-metallic transition of hydrogen a continuous transition?  Does the presence of a 10\% mixture of helium effect this transition?}
\item{What are the EOSs of hydrogen and helium along the internal adiabats of Jupiter and Saturn?}
\item{What is the temperature of hydrogen along the relatively ``cool'' adiabats of giant planets?}
\\
\\
\end{itemize} 

JJF acknowledges the support of an NASA Postdoctoral Program (NPP) fellowship and a travel grant from the HEDLA conference organizers.


\end{document}